**Experimental observation of thermal fluctuations in single superconducting Pb nanoparticles through tunneling measurements**

Ivan Brihuega[1,2] , Antonio M. García-García[3,4*], Pedro Ribeiro[4] , Miguel M. Ugeda[1,2], Christian H. Michaelis[1], Sangita Bose[1,5*]and Klaus Kern[1,6]

[1] Max Planck Institute for Solid State Research, Heisenbergstrasse 1, Stuttgart, D-70569, Germany.

[2] Univ. Autonoma Madrid, Dept. Fis. Mat. Condensada, E-28049 Madrid, Spain.

[3] T.C.M. Group, Cavendish Laboratory, University of Cambridge, J. J. Thomson Avenue, Cambridge CB3 0HE, United Kingdom

[4] CFIF, Instituto Superior Técnico, UTL, Av. Rovisco Pais, 1049-001 Lisboa, Portugal.

[5] Centre for Excellence in Basic Sciences**,** University of Mumbai, Vidhyanagari Campus, Mumbai –98, India.

[6] Institut de Physique de la Matière Condensée, Ecole Polytechnique Fédérale de Lausanne, CH-1015 Lausanne, Switzerland.

*An important question in the physics of superconducting nanostructures is the role of thermal fluctuations on superconductivity in the zero-dimensional limit. Here, we probe the evolution of superconductivity as a function of temperature and particle size in single, isolated Pb nanoparticles. Accurate determination of the size and shape of each nanoparticle makes our system a good model to quantitatively compare the experimental findings with theoretical predictions. In particular, we study the role of thermal fluctuations (TF) on the tunneling density of states (DOS) and the superconducting energy gap (Δ) in these nanoparticles. For the smallest*

[*] To whom correspondence should be addressed: amg73@cam.ac.uk, sangita.bose@gmail.com

*particles, $h \leq 13$nm, we clearly observe a finite energy gap beyond $T_c$ giving rise to a "critical region". We show explicitly through quantitative theoretical calculations that these deviations from mean-field predictions are caused by TF. Moreover, for $T << T_c$, where TF are negligible, and typical sizes below 20 nm, we show that $\Delta$ gradually decreases with reduction in particle size. This result is described by a theoretical model that includes finite size effects and zero temperature leading order corrections to the mean field formalism.*

## I. INTRODUCTION

Superconductivity in quantum confined systems has been a subject of research for the past few decades[1,2,3] . However, recent findings promoted by technological developments have revived the interest in this field [4,5,6,7,8,9,10,11,12,13,14,15,16,17,18]. These advances can shed light on the evolution of the ground state with particle size or the role of (thermodynamic) fluctuations on the stability of the superconducting state. Many earlier reports[19,20,21,22,23] have addressed some of these questions. However, a complete understanding of these problems is far from complete though some general features are broadly accepted.

The effect of thermal fluctuations (TF) on superconductivity in small particles has been probed previously in ensemble averaged nanoparticle systems through studies of specific heat and diamagnetism[24,25,26]. It is known that, as the dimension of the system is reduced below the superconducting coherence length ($\xi$), there are deviations from mean field behavior due to quantum and TF that lead to the smearing of the superconducting transition[27]. Interesting effects arise above the transition temperature ($T_c$) like the appearance of excess diamagnetism, conductivity, specific heat and tunneling currents. A “critical regime” can thus be defined where superconducting fluctuations dominate. In a zero dimensional superconductor where all dimensions are less than $\xi$, fluctuation effects lead to a large “critical regime” which in principle can be accessed experimentally. A detailed study of this “critical regime” is possible only through measurements on single, isolated superconducting nanoparticles. Moreover, a good knowledge of the geometry of the system is required to carry out a quantitative comparison between experiments and theory.

In this paper, we present an experimental study of the evolution of superconductivity in single, isolated nanoparticles as a function of size and temperature which overcomes these challenges.

Through our scanning tunneling microscopy (STM) measurements on individual Pb nanoparticles with sizes ranging between 3-30 nm grown *in situ*, in ultrahigh vacuum (UHV) conditions, we have addressed two fundamental questions: In the low temperature limit, how does superconductivity evolve as the particle size is decreased? For higher temperatures, how do TF affect the tunneling DOS and hence superconductivity? In order to answer quantitatively these questions we have compared the experimental results to the theoretical predictions of a model that includes both TF in the static path approximation (SPA)[28,29], finite size corrections to mean field and the leading corrections to the mean field formalism at zero temperature.

The paper is structured as follows. After this introduction, in Sec. II sample preparation and the experimental methodology to obtain the superconducting energy gap ($\Delta$) of the Pb nanoparticles are presented. In Sec. III A, we provide a theoretical description of the density of states (DOS) and the superconducting gap $\Delta$ based on the path integral formalism that describes the effect of thermal fluctuations at $T \sim T_c$. Sec. III B theoretically addresses the low temperature regime ($T \ll T_c$) including deviations from mean field predictions according to the Richardson formalism. The main results are presented and discussed in section IV, where the evolution of $\Delta$ with the particle size in the low temperature limit and the thermal fluctuations giving rise to a finite gap for $T > T_c$ are shown in IV A and IV B respectively. In section V, the validity of the Dynes expressions in the nanoscale regime and the role of the broadening parameter $\Gamma(T)$ are investigated. Finally the main conclusions of the present work are summarized in Sec. VI.

## II. EXPERIMENTAL METHODS

The experiments were performed in a UHV system (base pressure < $5 \times 10^{-11}$ Torr) combined with a home-built $^3$He low temperature STM. Pb isolated nanoparticles of 3-30 nm height were grown *in situ* on top of a BN/Rh(111) surface by means of buffer layer assisted growth (see Fig. 1a). BN, an ultrathin insulating layer with an electronic gap of 5-6 eV and a thickness of ~ 0.25 nm, electronically decouples the particles from the metallic Rh(111) substrate[30](For details of the preparation see Appendix A). Our high resolution STM images reveal that particles are hemispherical to a good approximation (See Appendix A). Differential conductance spectra (dI/dV *vs* V) were measured with a tungsten (W) tip on each nanoparticle with different heights using the lock-in technique (50 μV at 830 Hz voltage modulation). A stabilization current of 0.1 nA and an initial sample voltage of 8.0 mV was used to measure all the tunneling spectra. The additional energy broadening (other than thermal broadening) due to instrumental noise in our system was calibrated by measuring the tunneling conductance of a Pb(111) single crystal at 1.5K (open squares in Fig 1b). An upper limit to the contribution of the broadening of the spectra due to an additional instrumental noise of 20μeV at 1.5K was obtained (see Appendix B for details). The calibration of the sample temperature was performed by measuring the superconducting critical temperature for bulk Pb and the expected $T_c$ of 7.25 K was obtained (see Appendix B).

Fig. 1b shows unprocessed experimental dI/dV spectra measured at $T \ll T_c$ for different particle sizes which gives their density of states (DOS). We fitted the curves with the tunneling equation[31]:

$$G(V) = \left.\frac{dI}{dV}\right|_V = G_{nn} \frac{d}{dV}\left\{ \int_{-\infty}^{\infty} N_s(E,\Delta,\Gamma)\{ f(E) - f(E-eV)\} dE \right\} \qquad (1)$$

where, $N_s(E,\Delta,\Gamma)$ is the DOS of the superconductor, $f(E)$ is the Fermi-Dirac distribution function, $G_{nn}$ is the conductance of the tunnel junction for $V \gg \Delta/e$ and $\Gamma$ is an effective broadening parameter which in bulk superconductors is related to the quasiparticle lifetime[32,33].

As it will be shown in section V, in the case of nanoparticles, it can also have contributions from TF.

In Pb, a strong coupled superconductor, $N_s(E,\Delta,\Gamma)$ should in principle be obtained directly from the Eliashberg's mean-field formalism which accounts for recombination processes and electron-phonon scattering. However a simpler DOS ansatz was proposed by Dynes and coworkers[32]:

$$N_s(E,\Delta,\Gamma) = \mathrm{Re}\left[\frac{|E| + i\Gamma(T)}{\sqrt{(|E| + i\Gamma(T))^2 - \Delta(T)^2}}\right] \tag{2}$$

This is broadly used since the values of Δ and Γ thus obtained are in excellent agreement with the theoretical predictions of the Eliashberg's formalism. We used equations 1 and 2 to fit our experimental spectra with Δ and Γ as fitting parameters. As it can be seen in Fig. 1b, there is an excellent agreement between the experimental data and the theoretical fits giving unique values of Δ(T) and Γ(T) which characterizes ideally the superconducting state of each Pb nanoparticle.( see Appendix B for the description of the fitting procedure).

Solid symbols in Fig 1c show the size variation of the experimental superconducting gap Δ for low temperatures T=1.1-1.25K obtained from the fits using equations 1 and 2. We observe that for large particles (>20 nm), Δ is similar to that of bulk Pb (~1.36 meV) and subsequently decreases gradually as particle size is reduced.

## III. THEORETICAL ASPECTS OF SUPERCONDUCTIVITY IN A ZERO DIMENSIONAL SUPERCONDUCTOR

Although Dynes ansatz nicely reproduces the experimental conductance spectra (see Figs 1 and 2) it does not give any information about the physical phenomena relevant at these length scales.

One of the main goals of this paper is to overcome this limitation by providing a quantitative theoretical description of the experimental results. For that purpose we combine different theoretical tools, from the path integral formalism in the static path approximation (SPA) for $T \sim T_c$ to Richardson's equations that describe deviation from mean-field results in the low temperature limit. In the following we provide an introduction to these techniques.

**A. $T \sim T_c$ : Description of thermal fluctuations by the path integral formalism**

The path integral formalism in the so called static path approximation[28] (SPA) is a powerful tool to describe the interplay between superconductivity and thermal fluctuations in a zero dimensional nanoparticle. We note that for $T \ll T_c$ , corrections to mean field due to thermal fluctuations are small and others not included in SPA become relevant. Consequently other techniques, such as the Richardson formalism, must be employed in order to describe superconductivity beyond the mean-field limit in this region (see next section).

SPA assumes $\Delta$ to be space-time homogenous and can be used in our case as the system size is lower than the coherence length. Explicit analytical results are obtained for the DOS and $\Delta$ using this treatment as described below.

Superconductivity in the nanoparticle is modeled by the usual BCS Hamiltonian,

$$H = \sum_{\alpha} \varepsilon_\alpha \, c_\alpha^\dagger - \lambda\delta \sum_{\alpha,\alpha'>0} c_\alpha^\dagger \, c_{-\alpha}^\dagger \, c_{-\alpha'} c_{-\alpha}$$

where $\lambda$ is the coupling constant that describes the effective electron-phonon attraction, $\delta$ is the mean level spacing, $\varepsilon_\alpha$ are the eigenvalues of the one-body problem (for a free particle in almost hemispherical geometry). States labeled by $\alpha$ and $-\alpha$ are related by time reversal symmetry. By a Hubbard-Stratonovich transformation, the partition function of the system ( $Z = Tr\left[e^{-\beta H}\right]$ ) can be expressed in terms of a complex gap variable $\Delta(\tau, r)$. The spatial dependence of $\Delta(\tau, r) \approx$ $\Delta(\tau)$ is negligible since the coherence length of metallic particles is typically larger than the particle size. It is worth noting that the imaginary time dependence of the energy gap is related to quantum fluctuations (QF). We show in Sec. III B that for Pb this contribution is only relevant for particles with a typical size h < 5 nm. Therefore to a good approximation the complex gap variable is also homogenous in time. This amounts to completely neglect quantum fluctuations but keep thermal fluctuations in the partition function (Equation 3b). We determine Z explicitly which permits us to compute different observables related to the experimental results like the density of states (DOS) and the superconducting energy gap (Δ).

**A1. Density of states (DOS)**

The normalized density of states (DOS), $N_s(\omega)$ is related to the imaginary time Green's function $\mathcal{g}(\omega + i\Gamma, \alpha)$ by $N_s(\omega) = -\lim_{\Gamma\to 0^+} \frac{1}{\pi N} \sum_\alpha Im[\mathcal{g}(\omega + i\Gamma, \alpha)]$. In the SPA approach this Green's function is given by

$$\mathcal{g}(i\omega_n, \alpha) = \left(\frac{Z}{Z_0}\right)^{-1} \int_0^\infty d\Delta\, \Delta\, e^{-\beta\mathcal{A}(|\Delta|)} \mathcal{g}_D(i\omega_n, \alpha)$$

where $\mathcal{g}_D(i\omega_n, \alpha)$ is the usual superconducting Green's function. We simplify this expression to obtain the DOS given by:

$$N_s(\omega) = \left(\frac{Z}{Z_0}\right)^{-1} \int d\Delta\, \Delta\, e^{-\beta\mathcal{A}(|\Delta|)} \mathrm{Im}\left[\frac{\omega + i\Gamma}{\sqrt{\Delta^2 - (\omega + i\Gamma)^2}}\right] \qquad \text{3a}$$

$$\frac{Z}{Z_0} = \int d\Delta |\Delta| \mathrm{e}^{-\beta\mathcal{A}(|\Delta|)} \qquad \text{3b}$$

with $\mathcal{A}(|\Delta|) = \left\{(\lambda\delta)^{-1}|\Delta|^2 + \sum_\alpha(|\varepsilon_\alpha| - \xi_\alpha) - \frac{2}{\beta} log\left(\frac{e^{-\beta\xi_\alpha}+1}{e^{-\beta|\varepsilon_\alpha|}+1}\right)\right\}$, $\xi_\alpha = \sqrt{\varepsilon_\alpha^2 + \Delta^2}$, $\beta = 1/T$, δ is the mean level spacing and Γ is a broadening parameter which accounts for scattering or recombination processes, escape rates from nanoparticles, instrumental broadening etc. The sum

over $\alpha$ above is restricted to energies less than the Debye energy ($E_D$). $\varepsilon_\alpha$ denotes the energy levels of the one-body problem which in our case are the eigenvalues of the Schrodinger equation in a close to hemispherical particle.

In order to compare the experimental tunneling conductance spectra with the theoretical formalism of SPA, we calculate dI/dV or G(V) by substituting equation 3a in equation 1.

**A2. Superconducting energy gap**

Experimentally the superconducting energy gap is obtained by fitting the experimental G(V) with the theoretical expression given by equations 1 and 2 . This experimental gap is compared to the theoretical prediction within the SPA approach. We note first that since our model includes thermal fluctuations the gap is described by a certain distribution function. The moments of this distribution are related to the spectral gap measured in the experiments. When fluctuations become important there is a certain ambiguity in the definition of the gap as different definitions lead to the same gap in the bulk limit. However these differences should be small. Keeping this in mind, we have chosen the correlation function:

$$\Delta_{SPA}^2 = (\delta\lambda)^2 \sum_{\alpha,\alpha'>0} \langle c_\alpha^\dagger c_\alpha^\dagger\, c_{-\alpha'} c_{-\alpha}\rangle - \left[\langle c_\alpha^\dagger\, c_\alpha\rangle\langle c_{-\alpha}^\dagger\, c_{-\alpha}\rangle - \langle c_{-\alpha}^\dagger\, c_{\alpha'}\rangle\langle c_\alpha^\dagger\, c_{-\alpha'}\rangle\right]$$

Using the explicit expression for the SPA partition function (equation 3b) it is possible to rewrite the above expression as:

$$\Delta_{SPA}^2 = \left(\frac{Z}{Z_0}\right)^{-1} \int d\Delta\ e^{-\beta\mathcal{A}(|\Delta|)} |\Delta|^3 \left[\lambda \sum_\alpha \frac{\tanh\left(\beta\xi_\alpha/2\right)}{2\xi_\alpha}\right]^2 \qquad 4$$

Since thermal phonons are less effective to pair electrons, the electron-phonon interaction strength ($\lambda$) decreases as temperature increases. In order to simulate this effect, the value of $\lambda$ we use for each temperature is the result of a simple quadratic interpolation between the values of $\lambda$ at low temperature and those around the critical temperature. The values of $\lambda$ used in the interpolation were obtained by fitting the experimental dI/dV with the theoretical G(V) (equations 1 and 3a) obtained by the SPA approximation.

We will use theoretical expression 4 to obtain energy gaps from the SPA approach which are compared with the experimental energy gap (see figure 4).

**B. $T \ll T_c$: Deviations from mean field predictions and the Richardson formalism**

At low temperatures ($T \ll T_c$), QF of the order parameter and other types of finite size corrections not included in SPA (in this limit SPA is equivalent to mean-field) can become important. Therefore, to calculate the low temperature behavior of the superconducting nanoparticles we proceed in the following way. We start with the BCS gap equation for a hemispherical particle of height **h** which can be recovered by a zeroth order saddle point calculation of the partition function (equation 3b). This leads to:

$$\partial_{|\Delta|}\mathcal{A}(\Delta) = 0 \Rightarrow \lambda^{-1} = \sum_\alpha \frac{\tanh\left(\beta\xi_\alpha/2\right)}{2\xi_\alpha} \qquad 5$$

where the sum is again restricted to the interval $[-E_D, E_D]$ around the Fermi energy. This mean-field approach breaks down as the energy gap becomes comparable with the mean level spacing. In our experiments this occurs for h < 20nm.

In this region the Richardson formalism[34], valid at T=0, is a powerful tool to describe deviations from mean-field results. For instance it was found that, for $\delta/\Delta_0 <1$, (where $\Delta_0$ is the solution of Eq. 5 in the bulk limit) deviations in the energy gap due to QF are very small $\sim(\Delta/E_D)(\delta/\Delta)$ ($E_D$ is the Debye energy)[35,36]. The energy gap is therefore strongly modified by QF only for $\delta/\Delta > 1$. We can safely neglect the role of QF on $\Delta$ since $\delta/\Delta > 1$ corresponds to heights < 5 nm close to the minimum size experimentally studied here. The second important result from the Richardson's formalism is that the leading finite size correction ($\delta/\Delta_0$) to the mean field limit is a blocking effect that can be implemented by simply removing the two levels closest to Fermi level from the usual BCS gap equation[35]. This implies that for grains with an odd number of particles the state occupied by the unpaired one does not participate in pairing (See Appendix C for details)

For nanoparticles, $E_F$ and especially $E_D$ depend on the particle size. However in all our calculations we stick to the values of bulk Pb since its size dependence is relatively weak and poorly understood theoretically. Hence, a theoretical modeling of the size dependence of these quantities would involve the introduction of additional free parameters which would weaken the applicability of the theory. The resulting theoretical expression of the gap is valid up to corrections of the order $\sim(\delta/\Delta_0)^2$, which for Pb nanoparticles implies a size regime h > 5-6 nm. In order to mimic the effect of dynamical phonons present in the Eliashberg formalism used to describe a strong coupling superconductor, we employ size dependent electron-phonon interaction strength ($\lambda$). A simple quadratic interpolation scheme is employed using $\lambda$ obtained from fitting of the experimental G(V) at $T \ll T_c$ with the SPA G(V) given by equations 1 and 3a (see Appendix D for details). In addition, according to the experimental results, in Pb grains fluctuations of the energy gap, caused by shell effects in the spectral density, as a function of the particle size are small[14] in comparison with those of Sn. This is not surprising as inelastic scattering and other quantum decoherence effects, that shorten the coherence length, and consequently suppress shell effects[14,17], are enhanced in strongly coupled superconductors such as Pb. As a consequence of this, in our theoretical model we have smoothed out these fluctuations and have not taken into account other coherence effects such as the size dependence of the chemical potential and matrix elements.

Based on the above facts we calculate theoretically the particle size variation of Δ at low temperatures (Fig. 1c) by solving exactly the gap equation 5 for a close to hemispherical grain (this accounts for BCS mean-field finite size effects) but including the above mentioned blocking effect which account for the leading deviations from mean field.

## IV. RESULTS AND DISCUSSIONS

After the theoretical introduction and the description of the experimental methods we are now ready to compare theory and experiment. We start with the evolution of superconductivity with particle size in the low temperature limit.

### A. Evolution of the energy gap with the particle size in the low temperature limit

Fig 1c (solid symbols) shows the particle size variation of the superconducting energy gap (Δ) at low temperatures ($T << T_c$) obtained from fitting the experimental tunneling spectra with the Dynes ansatz (Eqs. 1 and 2). We observe that Pb particles larger than 20 nm show a superconducting gap similar to bulk Pb (~1.36 meV) which gradually decreases with reduction in particle size. As was explained in the previous section such gradual decrease of Δ with particle size can be modeled theoretically by considering finite size corrections within a mean-field approach and the leading corrections to the mean field formalism itself. For this purpose we use the theoretical treatment discussed in Sec. III B to calculate the particle size variation of Δ.

The black solid line in Fig 1c shows the theoretical prediction for the superconducting gap obtained for the electron-phonon interaction strength λ fixed to the Pb bulk value (λ=0.385). It nicely reproduces the decrease of Δ for diminishing sizes, in qualitative agreement with the experimental data. It is important to stress here that there are no input parameters included apart from the experimentally measured shape of the particles. The red line plot in Fig 1c shows how the small discrepancy between experiments and theoretical predictions vanishes when including $\lambda$ variation with particle (See Appendix D). Therefore, our results reveal that the observed evolution of $\Delta$ with particle size in Pb nanoparticles can be mostly explained just by invoking finite size effects.

**B. Thermal fluctuations and a finite gap for $T > T_c$**

For higher temperatures (1.3-8K) superconducting properties were also investigated by acquiring dI/dV for the Pb nanoparticles of different sizes at different temperatures. Figures 2 (a)-(c) show the tunneling spectra for two particles 23 and 10.5 nm high. Each spectrum is first fitted using the Dynes ansatz (Eqs. 1 and 2) giving the experimental $\Delta(T)$ and $\Gamma(T)$ as a function of temperature for each particle (Figs. 2b, d). From Figs. 2b, d, we observe that the evolution of $\Delta(T)$ is quite different for the two particles. While for the large particle it follows the characteristic BCS variation (black solid line), for the small particle clear deviations from BCS are visible at high temperatures. We also observe a finite energy gap beyond the mean field $T_c$ (where $T_c$ is defined as the temperature where the mean field BCS gap goes to zero). As was already discussed in the introduction we expect TF, controlled by the parameter $\delta/T_c$ , to induce deviations from mean field predictions. Therefore the "critical region" ($\propto (\delta/T_c)^{1/2}$) around $T_c$ becomes experimentally accessible for sufficiently small ($h \lesssim 13$ nm) Pb nanoparticles.

We note TF around $T_c$ are well described within SPA where only paths that are space and time independent are included in the partition function. The resulting theoretical expression for the energy gap from the SPA formalism is given by equation 4. Figures 3(a) and (b) show the experimental tunneling spectra for h = 10.5 nm and h = 8 nm respectively at different temperatures fitted with the theoretical expression given by SPA. SPA theory nicely reproduces the experimental data. Since TF are explicitly included in the SPA formalism, this implies that we identify the evolution of the tunneling DOS in small superconducting particles due to the influence of TF.

We would like to point out here that for the small Pb nanoparticles at temperatures close to and above $T_c$, the conductance varies by a very small amount (1-8%). Nevertheless, as shown in the insets of Figs. 2(a,c) and 3(a,b) the quality of the fits close to and above $T_c$ is still very good, allowing us to extract unique values of the experimental superconducting energy gaps for this crucial temperature window.

In Figure 4(a-d) we show the temperature evolution of the gap for four particle sizes (h = 23 nm, 13 nm, 10.5 nm and 8 nm). Here the symbols are obtained from the Dynes fits to the experimental tunneling spectra. The blue solid line is the theoretical expression using the SPA formalism (eq. 4). Again, we observe that while for the larger particles (h $\gtrsim$ 15 nm) the energy gap follows the expected BCS functional form (black solid line) for almost all temperatures, for the smaller ones the $\Delta(T)$ has a significant "tail" for $T>T_c$. Thus, we can conclude from our results that TF lead to a "fluctuation dominated regime" characterized by a finite energy gap beyond $T_c$ which persists up to a temperature T* ($>T_c$). Theoretically the energy gap should be finite even for higher temperatures. However its experimental detection is challenging as it becomes difficult to separate the signal from the background noise.

## V. RELIABILITY OF DYNES FITTING AND ANALYSIS OF THE BROADENING PARAMETER Γ(T) IN THE CASE OF NANOPARTICLES

In this last section we show the validity of the data analysis by Dynes expression in the nanoscale regime. We note that the agreement between theory and experiments found in the previous section is an indication that this is the case. This check is also important in order to support our claim that deviations from mean field predictions in the superconducting energy gap (Fig. 2 (b) and (d)) are caused by TF. We proceed by fitting the DOS in the SPA approach Eq. 3a (which accounts for TF) for different particle sizes with Dynes expression (see figure 5a). The good agreement between the energy gaps obtained from Dynes fits and SPA theoretical prediction Eq. 4 indicates that the Dynes' ansatz (Eq. 2) is suitable in the nanoscale region and captures the effect of TF. This also justifies our data analysis using the Dynes Ansatz in this region. In addition, such a good agreement also clearly confirms that the deviations from mean field results observed in experiments (Figs. 2(b) & (d) fitted using Dynes Ansatz) are caused by TF.

It is worth noting that in the Dynes expression the phenomenological parameter $\Gamma(T)$ which is the energy scale related to the phonon recombination and scattering and escape rates has also contributions from TF that peaks at $T_c$. However in the range of sizes we are investigating a quantitative analysis of TF by the experimental variation of $\Gamma(T)$ is much difficult to carry out than the one presented here for $\Delta(T)$.

Nevertheless, we can still show that in the case of nanoparticles $\Gamma(T)$ also contains information about the presence of TF. To this end, it is helpful to write $\Gamma_D$, the value of $\Gamma$ obtained by Dynes fitting $\Gamma(T) \equiv \Gamma_D = 1/\tau + \Gamma_{th}$ where $\tau$ is the finite lifetime of the quasiparticles and $\Gamma_{th}$ is the energy scale associated with TF. We note that $\Gamma_{th}$ is a finite size effect and not related to $\tau$. It is expected that the maximum contribution to $\Gamma_{th}$ due to TF should occur at $T_c$ where, according to the mean field prediction, the energy gap vanishes.

In bulk strongly coupled superconductors such as Pb the mean field Eliashberg's formalism provides a useful tool to compute $\tau$ due to scattering or recombination processes (see Eqs.

11,12,18 and Fig 7 of reference 33). For Pb good agreement between theory and experiment was found[32,37]. The total broadening in Pb increases with T. Around $T_c$ it is about $1/\tau \approx 0.1$meV.

In nanograins the possibility of escape from the particle provides an additional mechanism that reduces the lifetime. In addition a small instrumental broadening, also contributes to $\Gamma_D$. The substrate is an insulator but electrons can escape from the nanoparticles to the underlying metallic Rh substrate by either quantum tunneling or thermal activation. The former depends on the ratio of surface to volume and therefore will increase as the grain size is reduced. The latter will obviously increase with temperature. Although an accurate computation is not possible we expect that, in the range of sizes investigated here, the combined effect of all mechanisms contributing to quasiparticle lifetime (and thus to $\Gamma_D$ through $1/\tau$) will be $\gtrsim 0.2$ meV for $h < 20$nm. In order to make an estimation of the contribution of TF ($\Gamma_{th}$) to the total energy broadening $\Gamma_D$ we fit the theoretical SPA prediction of dI/dV by Dynes expression. Fig. 5 presents points generated by the SPA formula with a constant density of single particle states, for fixed $\lambda = 0.34$ and $\tau^{-1} = 0.4$meV and their fits by the Dynes expression for different values of T. The excellent agreement obtained for all temperatures is a clear indication that Dynes expression can still be used in the nanoscale regime. Fig. 5b shows the temperature dependence of $\Gamma_{th}$, extracted from the fits as $\Gamma_{th} = \Gamma_D - \tau^{-1}$, for different values of $\delta$. We observe a peak around the critical temperature which becomes broader as $\delta$ increases, corroborating that $\Gamma_{th}$ is associated with TF.

The scaling of $\Gamma_{th}$ with the mean level spacing, $\delta$ and hence particle size (see Fig. 5c) strongly depends on the temperature. $\Gamma_{th} \propto \delta$ for $T \ll T_c$. It is possible to show that this behavior agrees with the analytical prediction obtained considering higher order corrections to the mean field solution within the SPA approximation. As was expected, near the mean field $T_c$ thermal fluctuations becomes more important. It is possible to identify a region of size proportional to $T_c(\delta/T_c)^{\nu_\Gamma}$, $\nu_\Gamma < 1$, around $T_c$ where deviations from mean field results is not negligible.

In Fig. 5d the values of the fitted Dynes energy gap, $\Delta_D$, are compared with the theoretical prediction for the energy gap in the SPA approach Eq. 4 . The agreement is excellent. Moreover they were found to present the same finite size scaling behavior. These results are a further confirmation that the analysis of the experimental data by Dynes expression captures correctly the physics of TF. The lifetime of quasiparticles due to decoherence processes and interactions $1/\tau = 0.4$ meV was taken to be constant in this simple example. For a realistic model it should increase with T in a monotonic way i.e. it should not be sensitive to the phase transition.

## VI. CONCLUSIONS

To summarize, we report direct evidence of TF and the gradual breakdown of superconductivity in Pb nanoparticles as the size is reduced. The experimental data is well described by a theoretical model that includes TF, mean-field finite size effects and leading corrections to the mean-field formalism. TF give rise to a finite energy gap or “fluctuation dominated regime” around $T_c$. Our results are a first step to understand quantitatively the evolution of superconductivity with particle size and the role of thermal fluctuations for single small superconductors.

## APPENDIX

### A) Growth and Shape of the nanoparticles

The main goal of the present work is to address the superconducting properties of single isolated Pb nanoparticles. This requires the growth of nanoparticles on a substrate with no electronic states close to the Fermi level and to have these particles well separated from each other. Thus, the BN/Rh(111) system presenting an ultrathin insulating BN layer with a band gap of ~6 eV represents an ideal surface for the growth of the particles and for carrying out STM/STS experiments. To prepare the BN/Rh(111), we obtained a clean Rh(111) surface by repeated annealing and argon-ion sputtering cycles. The clean surface was then exposed to a 40 l (1 l=$10^{-6}$ Torr s) dose of Borazine gas with the substrate temperature held at 1070 K. This procedure leads to the formation of a BN insulating layer on top of the metallic substrate with a complete monolayer coverage[30].

Once the BN/Rh(111) surface was formed, nanometer sized Pb particles were deposited on top of it by using buffer layer assisted growth (BLAG), which is known to produce small particles with narrow size distribution . We first adsorbed a Xe buffer layer on the BN/Rh(111) surface at 50K, then evaporated Pb on top of the Xe, and finally desorbed the Xe layer by warming up the sample to room temperature. Pb nanoparticles form directly on the Xe buffer layer due to the reduced interaction with the substrate and grow in size during Xe desorption due to cluster coalescence until making contact with the surface. The final size of the nanoparticles can be tuned by adjusting the amount of deposited Pb and the buffer layer thickness. In order to study a wide range of nanoparticle sizes, we used two different sets of preparation parameters: 0.5 ML of Pb with 3,000 l Xe and 2.0 ML of Pb with 15,000l of Xe. This produced isolated Pb nanoparticles with heights between 3 and 30 nm.

The shape of these Pb nanoparticles has been considered to be approximately hemispherical. In order to verify this we developed a very simple model shown in Fig A1 (inset). The model assumes a hemispherical nanoparticle of radius *r* and a spherical tip apex of radius *R*. *Z* and $Z_0$ represent the tip-particle and tip-substrate tunneling distances respectively. According to this model, the ratio between apparent nanoparticle height (*H*) and the convoluted full width (*D*) is:

$$H/D = (r + z - z_0)/2 \cdot \sqrt{(R + r + z)^2 - (R + z_0)^2} \quad \text{(A1)}$$

if we assume that the tip and particle radii are much larger than the tunneling distances z and $z_0$ (~1nm), then $H = r + z - z_0 \cong r$ and (A1) can be simplified to:

$$\frac{H}{D} \cong \frac{r}{2} \cdot \sqrt{r(r + 2R)} \quad \text{(A2)}$$

This expression with only one free parameter (R) can be used to directly compare with experiments, as the observed STM height H is very closely related to the particle radius r. To prove the hemispherical shape of the nanoparticles we have fitted the ratio *H/D* (STM height/STM width) as a function of the nanoparticle height *H* as obtained experimentally from the STM topographic images. This comparison must be restricted to nanoparticles of sizes similar to tip apex radii (R~r), otherwise the tip convolution would hide any real feature of the nanoparticles. In figure A1 the ratio *H/D* is plotted for two different sets of nanoparticles corresponding to two large scale STM images (blue and red) in which the tip did not change during the scan. This ensures that all the nanoparticles in the particular topographic image are measured with the same tip radius *R* (R=30 nm for blue and R=15 nm for red). Both set of nanoparticles can be nicely fitted by means of the above expression (A2), confirming that the shape of the Pb nanoparticles is close to a hemisphere.

**B) Experimental resoultion and fitting procedure of the experimental conductance spectra by Dynes ansatz:**

The experimental output G(V), i.e dI/dV spectrum, was fitted by a standard algorithm that minimizes the square of the difference between the experimental data and the G(V) given by the Dynes ansatz (Equations 1 and 2 of main text). All points considered in the fitting carry the same weight. We obtain the experimental values for the energy gap $\Delta$ and the quasiparticle energy broadening $\Gamma_D$ from the fitting. As can be observed in Figure 2, Dynes fitting provides an excellent description of the experimental data. However an important point of the fitting procedure is described below.

The raw experimental data is rescaled so that G(V) goes to the unity for large voltage bias (V). For small particles h < 10-14nm and high $T \geq T_c$ we have observed that $\Gamma_D$, the value of $\Gamma$ obtained by Dynes fitting, is quite sensitive to small changes in the rescaling of G(V) and the fitting interval. This is not surprising as the dip in G(V) at V ~ 0 is quite weak with respect to the background noise. Therefore small errors in the rescaling have a strong impact in the value of the fitting parameters, specially $\Gamma_D$. In order to overcome this problem we have added a third fitting parameter that sets the rescaling of G(V). It is also necessary to carry out the fitting in a bias voltage interval which does not go much beyond the value of $\Delta$. We found that in most cases a 3 meV interval (from the origin) is a sensible compromise.

Broadening $\Gamma$ of the dI/dV spectra measured on the Pb superconducting nanoparticles is mainly due to intrinsic sources (scattering or recombination processes, thermal and quantum fluctuations, escape rates from nanoparticles) with minor contributions from extrinsic sources (the ac voltage modulation and instrumental noise). In order to calibrate the contribution to the broadening $\Gamma$ from the extrinsic sources and thus obtain an upper limit to the contribution to $\Gamma$ due to our instrumental noise, we have acquired dI/dV spectra with exactly the same tunneling parameters as for Pb nanoparticles on a bulk Pb sample. Open squares in Fig 1b correspond to a dI/dV spectra measured on bulk Pb at 1.5 K which has been fitted using the same broadened BCS DOS as the one used for the Pb nanoparticles (Dynes Ansatz, eq 1 and 2 from main manuscript). The best fit to the experimental data, which includes the 50 µV ac voltage modulation and the Fermi-Dirac broadening, gives a superconducting gap $\Delta$ = 1.36 meV, matching perfectly with the expected one for bulk Pb, and a broadening parameter $\Gamma$ = 10-20 µV. Therefore, we can estimate the maximum contribution due to our experimental noise to the broadening $\Gamma$ of our dI/dV spectra to be 20 µV. This value is one order of magnitude lower than the smallest $\Gamma$ that we have obtained on the Pb nanoparticles, which ensures that $\Gamma$ is mostly related to intrinsic sources in these nanoparticles. Moreover, our experimental system is very carefully shielded with respect to high-frequency (RF) voltage noise, which makes our effective temperature coincide (within 20 mK) with the actual temperature of the STM. The measurement of the superconducting critical temperature for bulk Pb was ultimately used to verify the calibration of the sample temperature. By measuring the evolution of $\Delta$ with temperature on a Pb bulk sample an expected $T_c$ value of 7.25 K was obtained.

We would like to point out that for the small Pb nanoparticles at temperatures close to and above $T_c$, the conductance varies by a very small amount (1-8%). Inspite of this, the quality of the fits to the spectra measured for the Pb nanoparticles with the Dynes expression close to and above $T_c$ is very good. For each spectrum, we use the least square fitting routine to minimize the $\chi^2$ to obtain $\Delta$ and $\Gamma$ for the best fit. We also take particular care to determine the uniqueness of the fits. As a representative, we show below the fitting done for the spectra measured on the 10.5 nm particle of figure 2(c) at T = 5.74 K. It is worth noting that in this spectrum the conductance varies by only 2%, since it is measured well above $T_c$ ($T_c$ =5.0 K for such a particle). Minimizing $\chi^2$ we obtain the best fit for $\Delta$= 0.21 meV and $\Gamma$ = 0.38 meV. To test the uniqueness of the fit, we change $\Gamma$(within a range) to compensate for the change in $\Delta$. It can be seen that the curve is reasonably fitted only with values of $\Delta$ and $\Gamma$ in a particular window of values. (see Fig A2. This gives us the error bars in $\Delta$ and $\Gamma$ which are shown in figures 2(b) and 2(d).

**C) Finite size effects included in the theoretical formalism (For $T << T_c$)**

Two distinct types of finite size effects in the one particle density of states were considered in the comparison with experimental results.

*Spectral fluctuations*

The first type of finite size corrections are simply mean effects related to the fact that, in superconducting nanoparticles, the spectral density of the one body problem is not constant in the interacting region around the Fermi level. A complete analytical treatment of these corrections, valid for any particle shape, was recently developed[17]. In practical terms these deviations are taken into account by considering explicitly the spectrum of the one body problem. In our case the particle can be modeled as a spherical cap of height h and radius R with R ~ h. We compute the spectrum numerically for a given ratio of h/R. In the hemispherical case, h/R=1, the eigenvalues are simply the roots of a Bessel function. For other ratios, we use a method based on a perturbative expansion[38] around the hemispherical geometry which is valid for $1-h/R << 1$.

*Blocking effects*

As the particle size is reduced to the point that $\delta$ becomes comparable to the bulk gap, finite size effects induce deviations from mean-field predictions. In the limit of vanishing temperatures the solution of the Richardson's equations[34,39] provide an exact account of the ground state and low energy excitations of the usual BCS Hamiltonian. Recently it was shown[35], from the exact Richardson's equations, that the leading correction $\delta/\Delta$ to mean-field has a simple interpretation within the mean-field formalism: it is equivalent to remove the two levels closest to the Fermi energy in the usual BCS gap equation (for a previous derivation of this result by using path integral methods see Ref. 36). Physically this is expected as the gap can be defined as the minimum energy to break a pair. In a finite system, as a result of the pair breaking, two electrons will occupy levels around the Fermi energy giving rise to a blocking effect in the sense that no pairing is possible at these energies. Using this we include the leading correction to mean field by simply removing the two levels closest to the Fermi energy in the definition of the action of our system. It was also shown in Ref. 35 that for $\delta/\Delta < 1$, quantum fluctuations, not included in our formalism, are at most of order $\sim(\Delta/E_D)(\delta/\Delta)$ which extends the applicability of our model until particle sizes h ~ 6 nm very close to the smallest particles which can be studied experimentally.

**D) Size dependence of the electron-phonon interaction parameter ($\lambda$)**

In our model $\lambda$ is an effective parameter that describes the strength of the interactions that lead to the superconducting state. It is well documented that in strongly coupled superconductors, such as Pb, $\lambda$ decreases with temperature as thermal phonons are less effective to glue electrons together. The dependence of $\lambda$ on grain size is less clear as in the nanoscale region several competing effects must be considered. Coulomb interactions and quantization effects in the phonon spectrum increase as the grain size is reduced. As a result we expect the effective coupling constant to decrease accordingly. On the other hand the increasing contribution of surface phonons as system size is reduced is expected to increase λ.

In order to make quantitative comparisons between theory and experiment it is thus necessary to employ a size dependent coupling constant. It is possible to estimate this size dependence by fitting the experimental differential conductance with the theoretical prediction from SPA approach where λ is a fitting parameter. In figure 1c we show the results of the fitting in the low temperature limit T=1-1.25K. We observe that for the largest grains the value of the coupling constant leads to an energy gap very close to the bulk one ~ 1.35meV. As a general rule the coupling constant decreases with the system size (Fig A3). However the size dependence for $h >$ 6-7nm is relatively weak $\leq 5\%$. It is tempting to speculate that the flattening observed for $h \sim$ 10nm is due to the interplay between surface phonons effects that enhance pairing and the rest of effects that tend to weaken it. We note (see Fig 1c) that quantitative agreement between theory and experiment is only achieved after this small size dependence of λ is included in the theoretical model.

**Figure Captions**

**Figure 1** (**a)** (Color online): The top panel shows a schematic of the isolated Pb nanoparticles on the BN/Rh(111) substrate. The lower panel is a typical STM image (162 X 125 $nm^2$) showing the general morphology of the isolated Pb nanoparticles of different sizes (3D plot obtained with the WSxM[40]) taken at a bias voltage of 1V and a tunneling current of 0.1nA., (**b)** Conductance spectra for particles with different sizes and T =1.1-1.25 K. The symbols denote the experimental raw data. The solid lines are the best fitting using Eq. 1 with the DOS $N_s$ given by the Dynes expression Eq. 2. Open squares correspond to a reference spectra measured in a Pb single crystal at 1.5 K **(c)** Experimental (blue circles) and theoretical (solid lines) of the average superconducting energy gap for T=1.2K as a function of particle size (h). In the black theoretical curve it was assumed that the electron-phonon coupling $\lambda$ = 0.385 does not depend on the particle size. In the red theoretical curve a size dependent $\lambda$(h) is taken (See appendix D and figure A3).

**Figure 2** (Color online): Conductance spectra *vs* temperature (T) for two particles of heights **(a)** 23 nm **(c)** and 10.5 nm. Experimental raw data are shown by open circles. Data in ash is taken at a temperature where no superconducting signal is obtained. The solid lines are the fits using Eq. 1. and 2. For clarity, the **inset** in figures 2(a)&(c) show the excellent fit to the conductance spectra at temperatures close to $T_c$ where the signal changes by 2-8% from the background. **(b)-(d)** show the variation of $\Delta$(T) (red solid circles) with temperature (T) for the two particles as obtained from the fits. Solid lines correspond to the variation expected from BCS theory.

**Figure 3**(Color online): **(a)** Conductance spectra for particle with height h=10.5 nm at different temperatures between T = 1.0-6.5K. The symbols denote the experimental raw data. The solid

lines are the best fits using Eq. 1 with the DOS $N_s$ given by the SPA expression Eq. 3a. **(b)** The same for h=8 nm. The **insets** in the two show a magnified scale where spectra close to and greater than $T_c$ have been plotted. The clear signal above the background at these temperatures clearly demonstrates that the energy gap does not go to zero at $T_c$ and we see the effect of fluctuations on the tunneling DOS. For clarity we show the spectra only for positive voltage.

**Figure 4** (Color online): Superconducting gap $\Delta(T)$ *vs* temperature T for different nanoparticles. Red circles show the experimental gap obtained from Eq.(1) with the DOS given by Eq.(2). Solid black lines correspond to the variation expected from BCS theory. Solid blue lines show the theoretical prediction which includes the effect of thermal fluctuations within the static path approach (see Eq.(4)). **(a)** h = 23 nm **(b)** h = 13 nm **(c)** h = 10.5 nm **(d)** h = 8 nm.

**Figure 5** (Colour Online): **(a)** Fit of the Dynes expression (red curves) to the dI/dV SPA prediction (open circles) for $\lambda$=0.34, $\delta$=0.02 meV and for different values of the temperature, **(b)** $\Gamma_{th}$, which quantifies thermal fluctuations, as a function of temperature for different values of $\delta$. Thermal fluctuations clearly increase as $\delta$ increases (i.e. as particle size decreases). The typical width of the peak provides an estimation of the region for which deviations from mean field results are more important, **(c)** The exponent $\nu_\Gamma$ for small $\delta$, describes the dependence of $\Gamma_{th}$ on $\delta$ through the expression, $\Gamma_{th}{\sim}\delta^{\nu_\Gamma}$ . The change observed around $T_c$ indicates a qualitative enhancement of the thermal fluctuations in this region, **(d)** Comparison between the SPA prediction (red curves) for the superconducting energy gap and those obtained by using Dynes expression to fit the dI/dV obtained by the SPA approach (open circles) for different particle sizes (different $\delta$).

**Figure A1**(Colour online): Plot of the experimental STM height/STM width as a function of the nanoparticle height h for two different tips (red and blue). Solid circles denote the experimental data. Solid lines are the fits obtained using Eq. (A2). The inset shows a schematic of the model used (see text).

**Figure A2** (Colour online): Normalized dI/dV spectrum measured on Pb nanoparticle of height 10.5 nm at 5.74K (black open circles). The fits to the data (solid lines) has been done using the Dynes expression (see main text) where the different lines are obtained with different fitting parameters. Best fit was obtained for $\Delta$ = 0.21meV and $\Gamma$ =0.38mV with a $\chi^2$ = 0.002 (green line).

**Figure A3** (Colour online): $\lambda$ as a function of the particle size (h). This result was obtained by fitting the experimental differential conductance by theoretical prediction of the SPA approach where $\lambda$ is a fitting parameter. The particle is close to hemispherical h ~ R.

**Figure1**

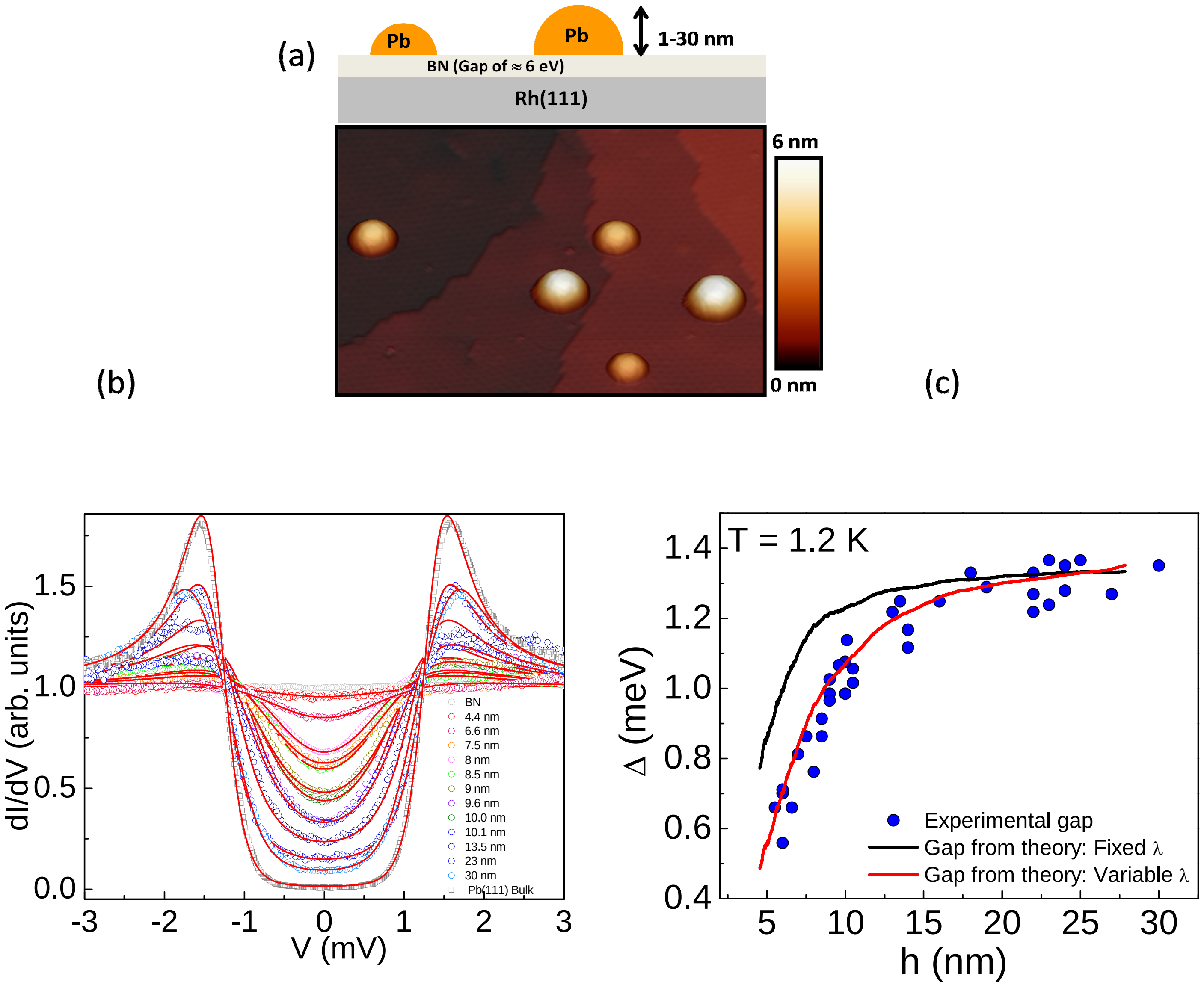

**Figure 2**

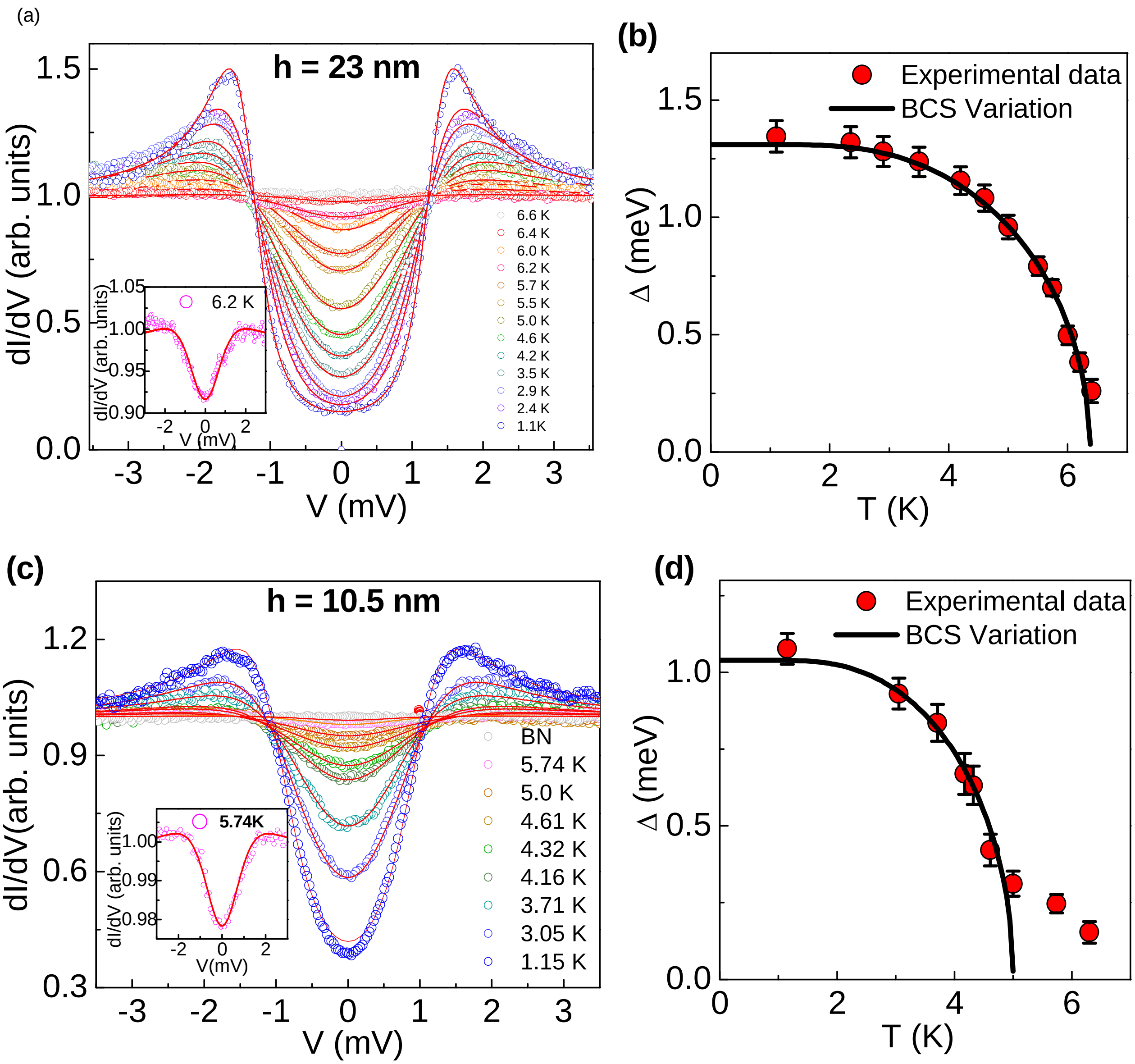

**Figure 3**

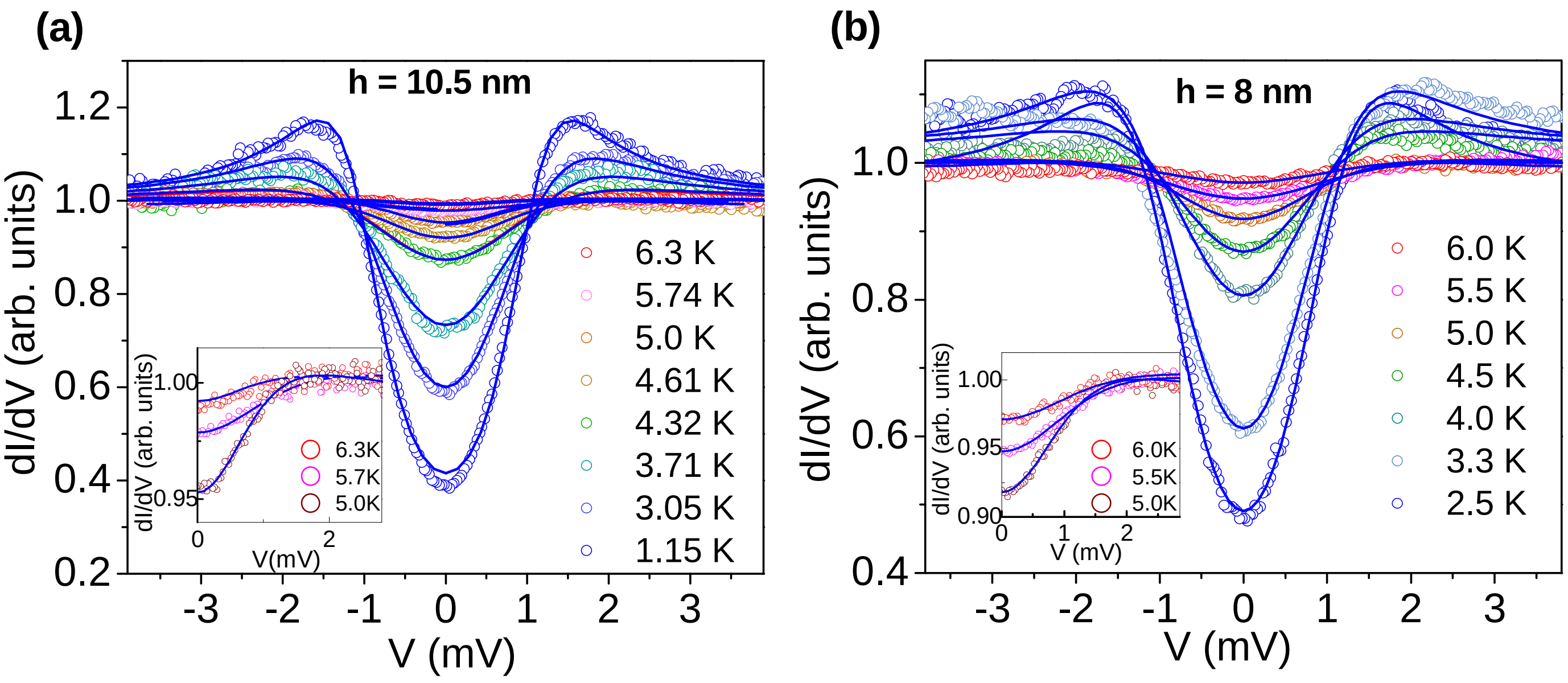

**Figure 4**

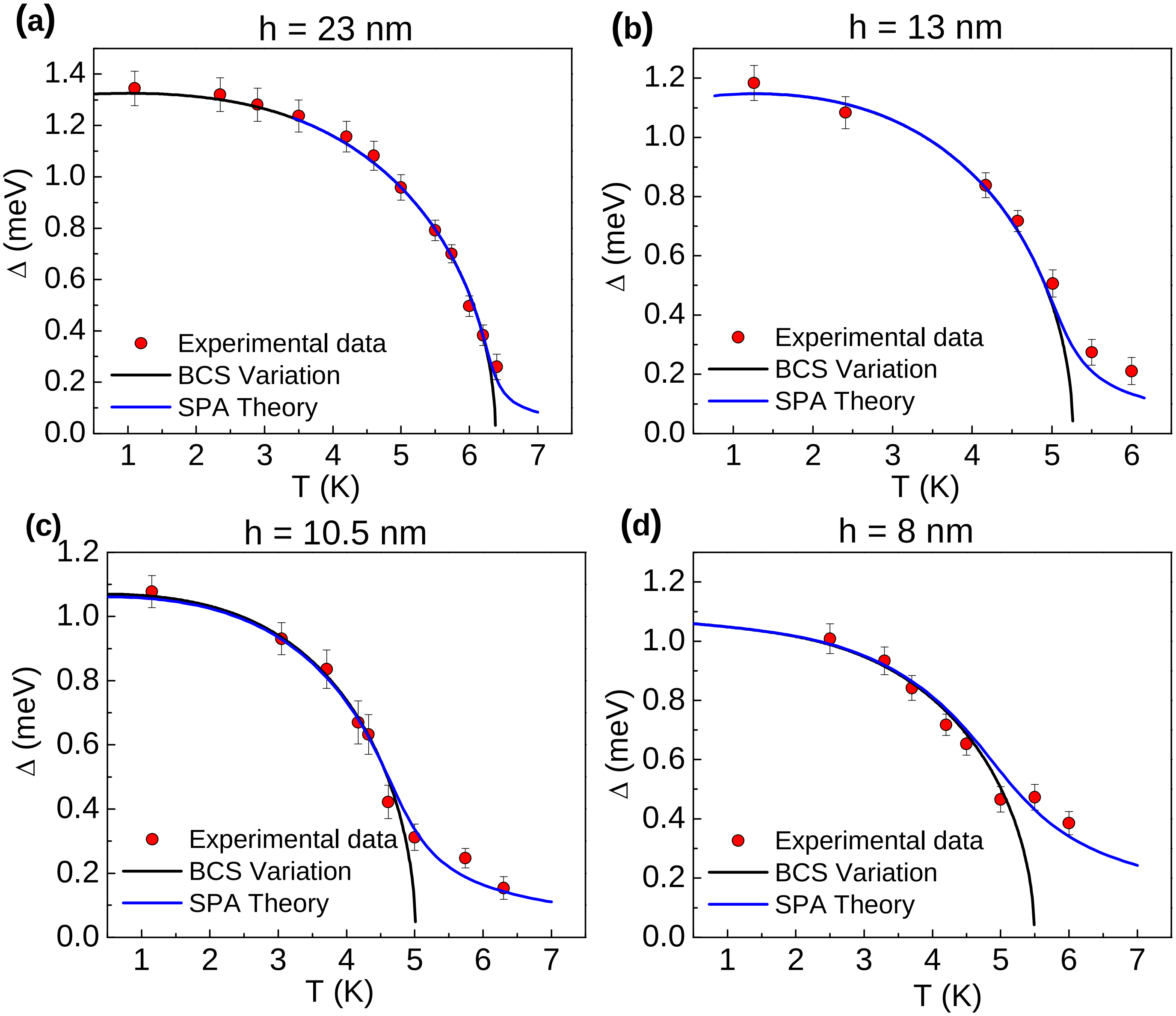

**Figure 5**

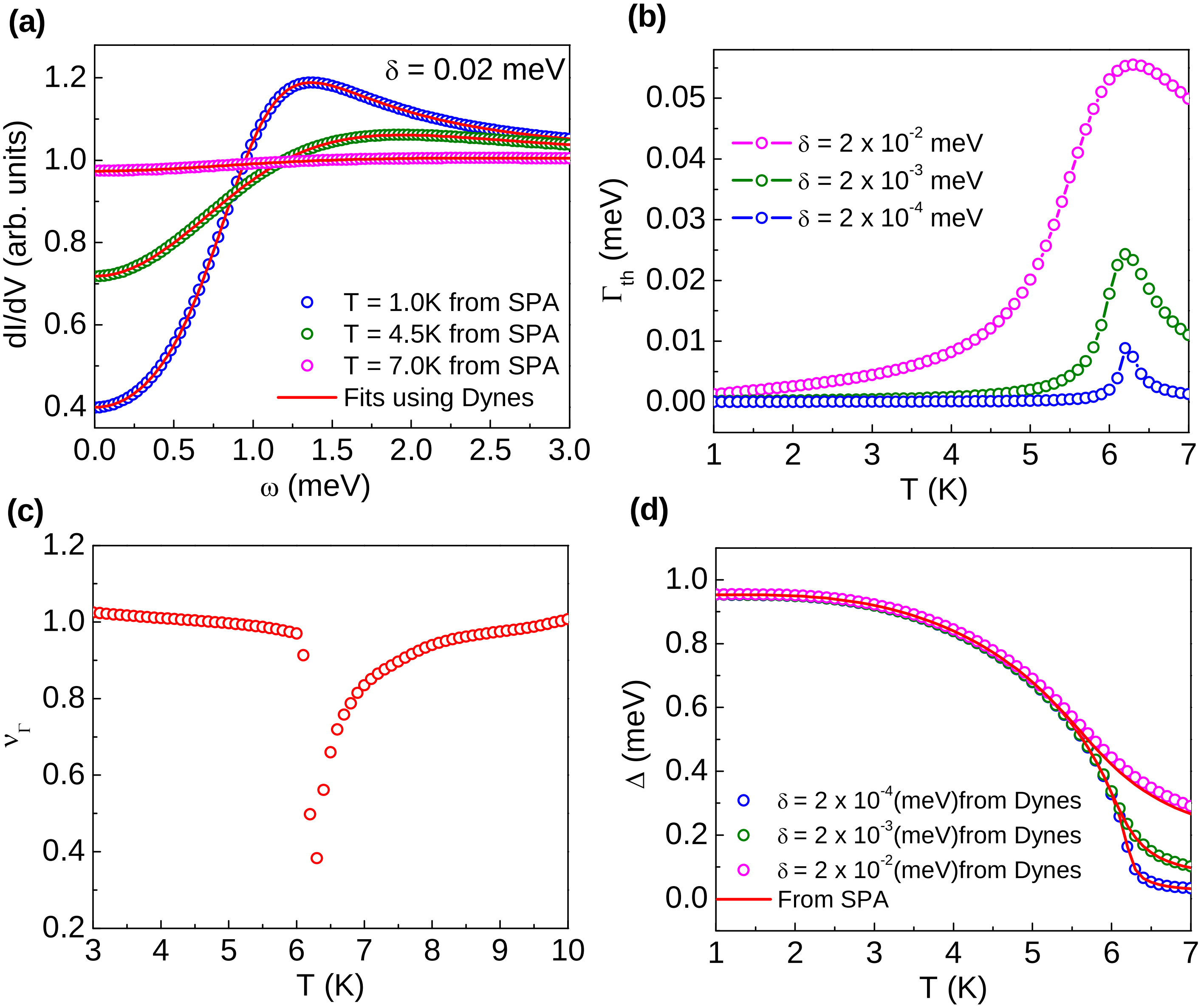

(a)
δ = 0.02 meV
dI/dV (arb. units)
ω (meV)
T = 1.0K from SPA
T = 4.5K from SPA
T = 7.0K from SPA
Fits using Dynes
(b)
Γ_th (meV)
T (K)
δ = 2 x 10^-2 meV
δ = 2 x 10^-3 meV
δ = 2 x 10^-4 meV
(c)
ν_Γ
T (K)
(d)
Δ (meV)
T (K)
δ = 2 x 10^-4(meV)from Dynes
δ = 2 x 10^-3(meV)from Dynes
δ = 2 x 10^-2(meV)from Dynes
From SPA

**Figure A1**

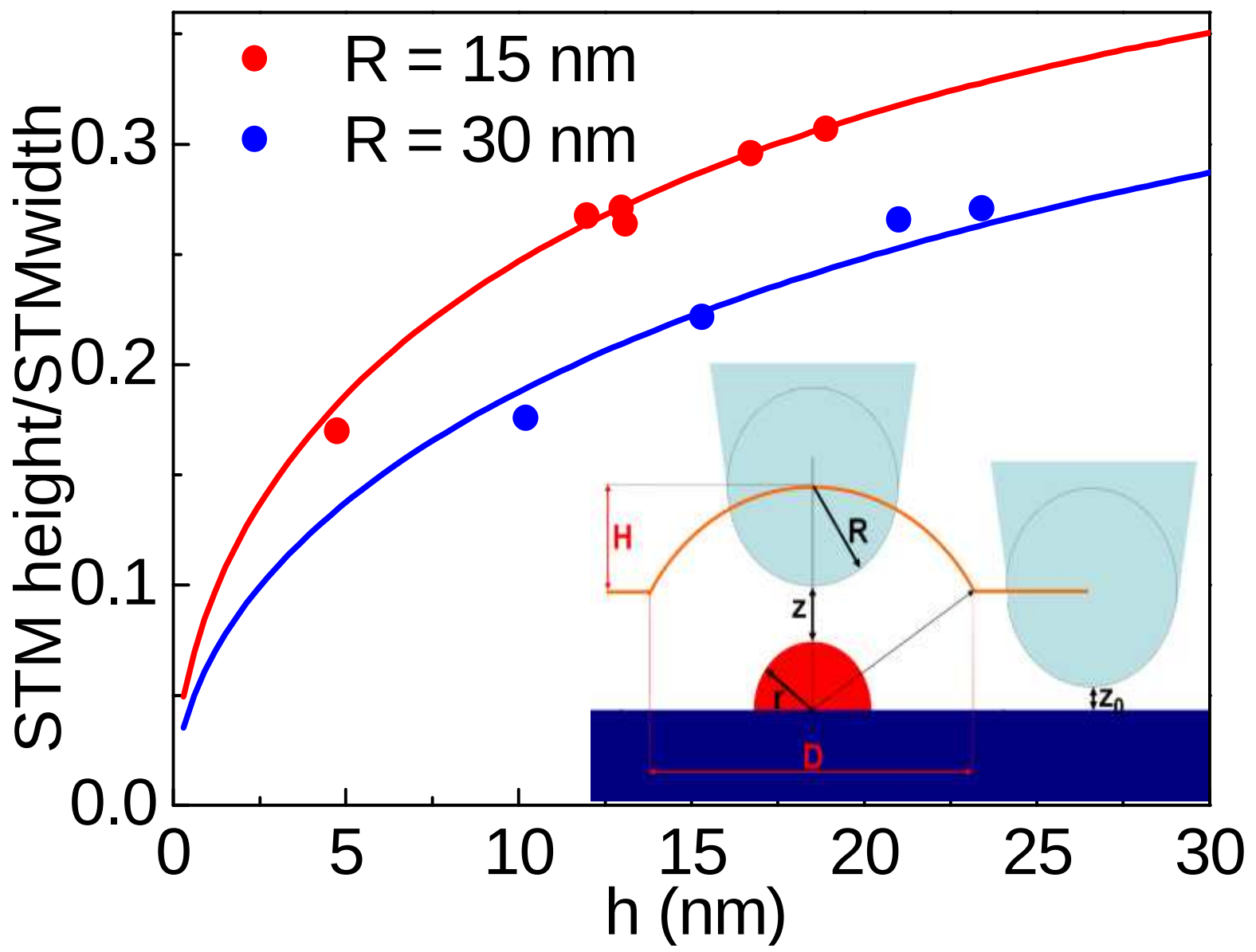

**Figure A2**

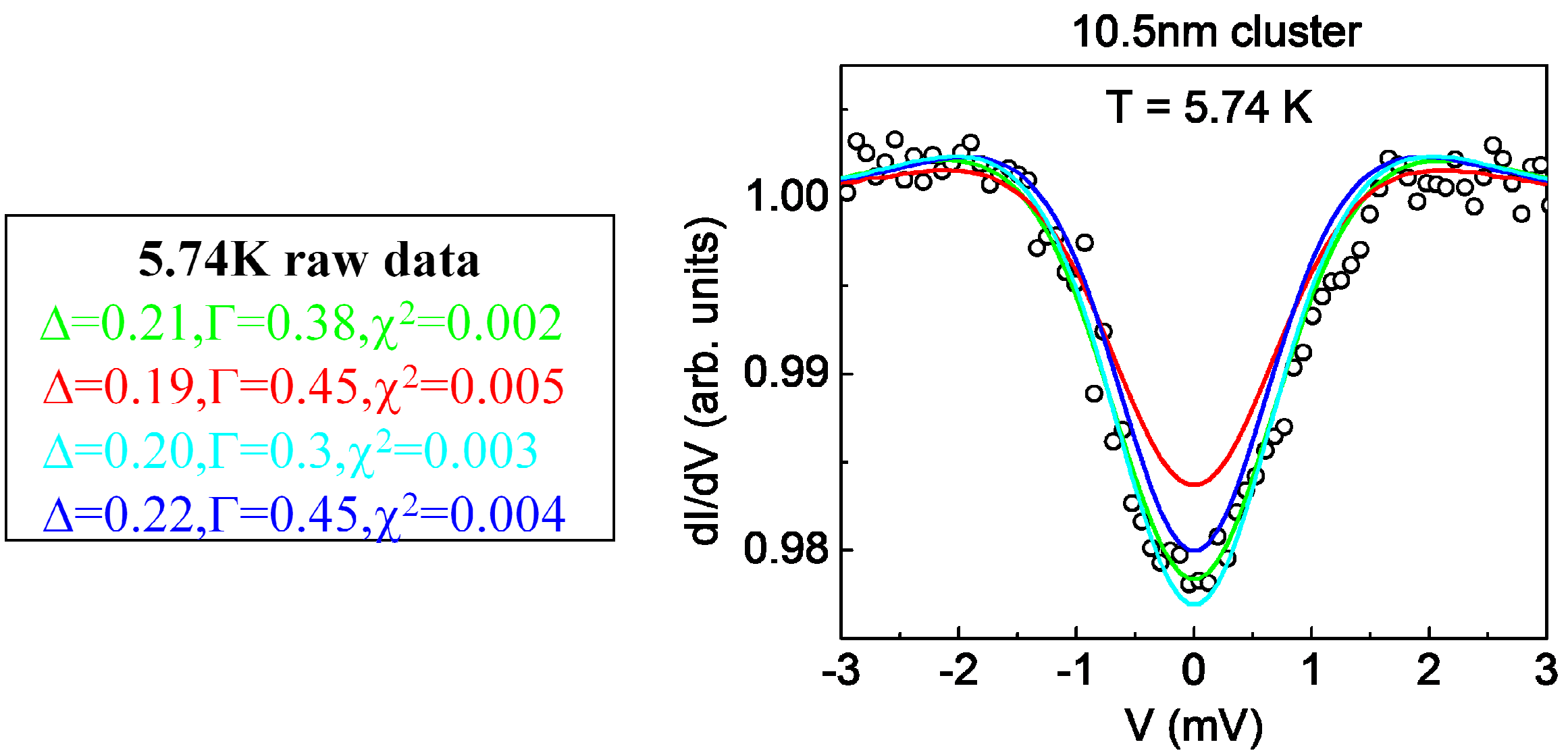

**Figure A3**

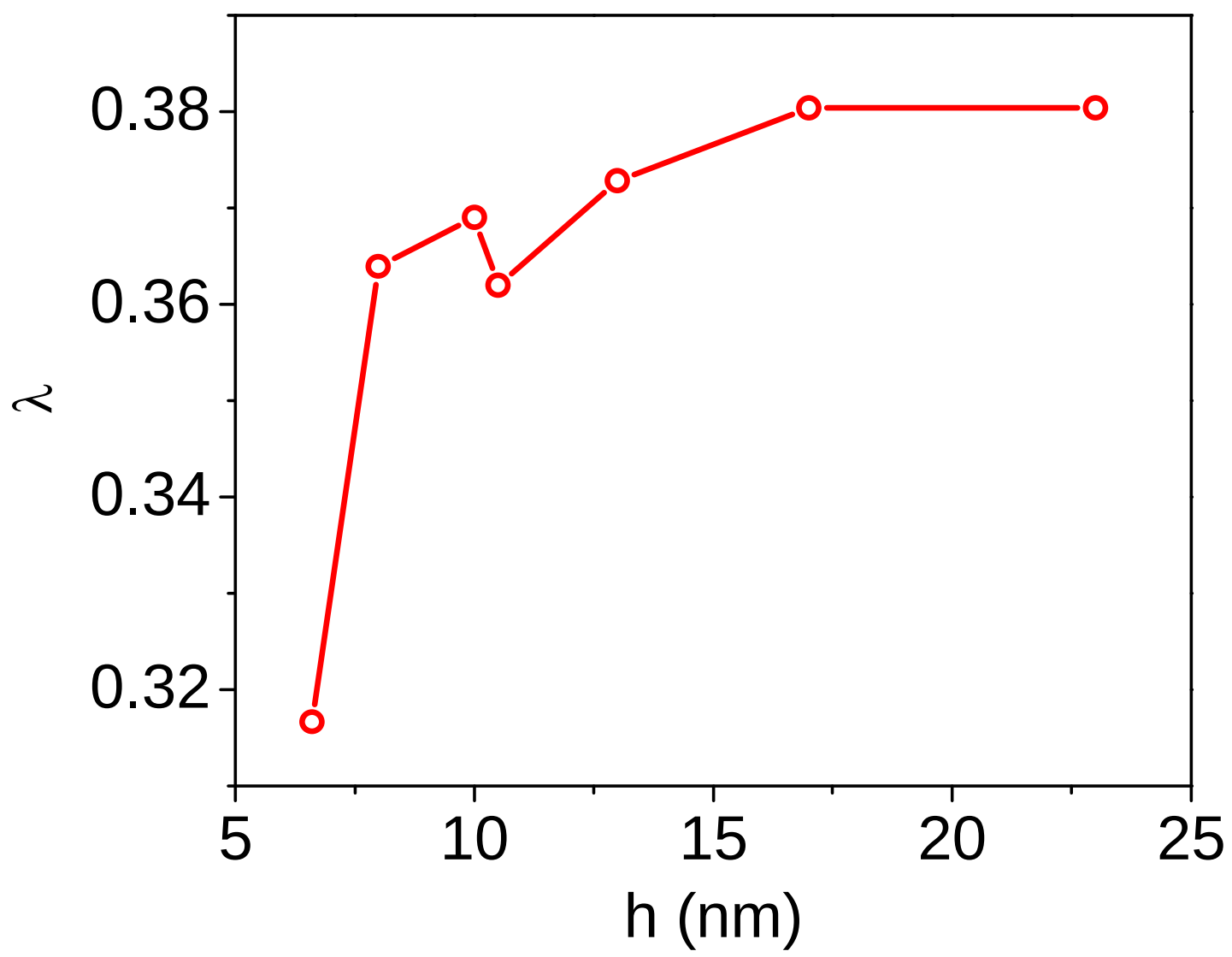

## References

[1] B. Abeles, R. W. Cohen, and G. W. Cullen, Phys. Rev. Lett. **17**, 632 (1966).

[2] M. Strongin, O. F. Kammerer, J. E. Crow *et al*., Phys. Rev. Lett. **21**, 1320 (1968).

[3] I. Giaever and H. R. Zeller, Phys. Rev. Lett. **20**, 1504 (1968).

[4] D. C. Ralph, C. T. Black, and M. Tinkham, Phys. Rev. Lett. **74**, 3241(1995).

[5] A. Bezryadin, C. N. Lau, and M. Tinkham, Nature **404**, 971 (2000).

[6] A. A. Shanenko *et al*, Europhys. Lett. **76**, 498 (2006).

[7] T. Nishio, M. Ono, T. Eguchi *et al.*, App. Phys. Lett. **88**, 113115 (2006).

[8] T. Nishio, T. An, A. Nomura et al., Phys. Rev. Lett. **101** (16), 167001 (2008).

[9] I. Guillamon *et al*. Nature Phys., **5**, 651 (2009).

[10] C. Brun *et al*., Phys. Rev. Lett. **102**, 207002 (2009).

[11] T. Cren, D. Fokin, F. Debontridder et al., Phys. Rev. Lett. **102**, 127005 (2009).

[12] K. Wang, X. Zhang, M. M. T. Loy *et al.*, Phys. Rev. Lett. **102**, 076801 (2009).

[13] T. Zhang *et al*., Nature Phys. **6**, 104 (2010).

[14] S. Bose *et al*., Nature Mat., **9**, 550 (2010).

[15] J. von Delft, Annalen Der Physik **10**, 219 (2001).

[16] V. Z. Kresin and Y. N. Ovchinnikov, Phys Rev. B **74**, 024514 (2006).

[17] A. M. García-García *et al.* Phys. Rev. Lett. **100**, 187001 (2008).

[18] D. Innocenti *et al*., Phys. Rev. B **82**, 184528 (2010).

[19] P.W. Anderson, J. Phys. Chem. Solids **11**, 26 (1959).

[20] W.-H. Li, C. C. Yang, F. C. Tsao and K. C. Lee, Phys. Rev. B **68**, 184507 (2003).

[21] S. Reich, G. Leitus, R. Popovitz-Biro and M. Schechter, Phys. Rev. Lett. **91**, 147001(2003).

[22] S. Bose *et al.*, Phys. Rev. Lett. **95**, 147003 (2005).

[23] S. Bose *et al.*, J. Phys. Condens. Matter **21**, 205702 (2009).

[24] T. Tsuboi and T. Suzuki, J. Phys. Soc. Jpn. **42**, 437 (1977).

[25] R. A. Buhrman and W. P. Halperin, Phys. Rev. Lett. **30**, 692 (1973).

[26] E. Bernardi *et al.*, Phys. Rev. B **74**, 134509 (2006).

[27] W. J. Skocpol and M. Tinkham, Rep. Prog. Phys. **38**, 1049 (1975).

[28] B. Mühlschlegel, D. J. Scalapino, and R. Denton, Phys. Rev. B **6**, 1767 (1972).

[29] R. Denton, B. Mühlschlegel, and D. J. Scalapino, Phys. Rev. B **7**, 3589 (1973).

[30] I. Brihuega *et al.*, Surface Science **602**, L95 (2008).

[31] M. Tinkham, *Introduction of Superconductivity* 2nd edn (McGraw-Hill, 1996).

[32] R. C. Dynes, V. Narayanamurti, and J. P. Garno, Phys. Rev. Lett. **41**, 1509 (1978).

[33] S. B. Kaplan *et al.*, Phys. Rev. B **14**, 4854 (1976).

[34] R. W. Richardson, J. Math. Phys. **18**, 1802 (1977).

[35] E. A. Yuzbashyan, A. A. Baytin, and B. L. Altshuler, Phys. Rev. B **71**, 094505 (2005).

[36] K. A. Matveev and A. I. Larkin, Phys. Rev. Lett. **78**, 3749 (1997).

[37] P. Hu, *et al.* Phys. Rev. Lett. **38**, 361 (1977)

[38] A.H, Rodriguez, C. Trallero-Giner, S.E. Ulloa, J. Marin-Antuna, Phys. Rev. B **63**, 125319 (2001).

[39] G. Sierra *et al.*, Phys.Rev. B **61** (2000) 11890

[40] I. Horcas, *et al*. Rev. Sci. Inst. **78**, 013705 (2007).